# Structural, electrical, magnetic and thermal studies of Cr-doped $La_{0.7}Ca_{0.3}Mn_{1-x}Cr_xO_3$ (0 ≤ x ≤ 1) manganites

Neeraj Kumar, H. Kishan and V. P. S. Awana[*]

Superconductivity and Cryogenics Division (CSIR), National Physical Laboratory (CSIR)

Dr. K.S. Krishnan Marg, New Delhi-110012, India

**Abstract**

We report detailed structural, electrical, magnetic and specific heat studies on $La_{0.7}Ca_{0.3}Mn_{1-x}Cr_xO_3$ manganites. Rietveld analysis of fitted and observed XRD patterns exhibited the single-phase nature of all the studied materials, which crystallize in *Pbnm* space group. Successive substitution of Cr at Mn site in $La_{0.7}Ca_{0.3}Mn_{1-x}Cr_xO_3$ manganites increases the electrical resistivity and decrease the characteristic insulator-metal transition temperature ($T_{IM}$) of the parent compound along with a hump like feature for higher Cr-content (x > 0.06) samples. The hump structure basically signifies the onset of anti-ferromagnetic (AFM) interactions as inferred by both the magnetic and infra-red (IR) spectroscopy studies. The systematic suppression of FM state results in a spin glass (SG) like behavior. IR studies revealed that the vibration mode at 413 $cm^{-1}$ being associated with internal bending of $MnO_6$ octehedra, becomes softer, indicating an increase in distortion and hence the possible spin glass behavior. The critical exponents (α, β and γ) are calculated from the heat capacity ($C_P$) data near the $T_{IM}/T_{FM}$. The same exhibited variations of their values with doping. In particular the value of β increases from 0.37(x = 0.0) to 0.43(x = 0.04), clearly indicating the coexistence of both long and short range magnetic orders i.e. tendency towards SG state for Cr doped samples. On the basis of present results, it is suggested that Cr dilutes double-exchange (DE) based FM and rather promotes the AFM based super exchange interactions (SE) via $Cr^{3+}/Mn^{4+}$ ions. Substitution of Cr systematically destroys both the metallic state and long range ferromagnetic (FM) order.

PACS: 75.47.Lx, 71.38.Ht, 71.30.+h

*Keywords*: Manganites, Insulator-Metal Transition, Heat Capacity, and Magnetization.

*Corresponding author's address: Dr. V.P.S. Awana
National Physical Laboratory, Dr. K.S. Krishnan Marg, New Delhi-110012, India
Fax No. 0091-11-45609310: Phone No. 0091-11-45609210
e-mail-awana@mail.nplindia.ernet.in; www.freewebs.com/vpsawana/

# 1. INTRODUCTION

Magneto-resistance (MR) is the relative change in the electrical resistivity of any material upon the application of magnetic field, defined as

$$MR = [(\rho_H - \rho_0)/(\rho_H)]*100.$$

This is one of the important parameters for magnetic data storage, magnetic sensors and spintronic devices [1]. During last decade, very high value of MR ($\sim 10^6\%$) was reported in the perovskite manganites materials and is known as colossal magnetoresistance (CMR) [1, 2]. In the perovskite structure ($ABO_3$ type), A is generally trivalent rare-earth ion like $La^{3+}$, $Pr^{3+}$, $Nd^{3+}$ or divalent alkali-earth ion like $Ca^{2+}$, $Sr^{2+}$, $Ba^{2+}$ etc. whereas B is the manganese ion, which can also be replaced by other trivalent transition metal ions. The parent compounds like $LaMnO_3$, $PrMnO_3$ etc. are antiferromagnetic insulators (AFMI) [1, 2]. However, with strategic substitution of divalent ions at A site, some of the $Mn^{3+}$ ions ($3d^3$, $t_{2g}^3 e_g^1$) convert into the $Mn^{4+}$ ions ($3d^3$, $t_{2g}^3 e_g^0$) resulting in fascinating physical phenomena like paramagnetic (PM) insulator to ferromagnetic (FM) conductor or various mixed magnetic phases viz. canted anti-ferromagnetic (AFM)/Spin Glass (SG) coupled with charge/orbital ordered states in a particular doping range [3]. The conduction in manganites has been proposed on the basis of the double exchange mechanism between $Mn^{3+}/Mn^{4+}$ via oxygen along with the strong electron-phonon coupling [4]. In manganite materials, the CMR effect is observed near the insulator-metal (I-M) transition, which can be tuned by changing the chemical composition of the compound. Doping at the rare-earth site indirectly affects the conduction mechanism with its repercussion on bandwidth and bond angle between the manganese ions [5]. On the other hand doping at the manganese site directly affects the conduction mechanism and hence one can tailor the properties of perovskite manganites in more effective way. Therefore, doping at Mn-site with other elements is expected to provide some important clues concerning the conduction mechanism of CMR as well as possible enhancement/ change in the MR of manganites materials. It was found that doping at Mn-site generally results in the decrease of the I-M transition temperature [6, 7]. Maignan et al. reported nearly invariant $T_c$ (Curie temperature) values in Cr-doped $Sm_{0.56}Sr_{0.44}Mn_{1-x}Cr_xO_3$ system [8]. On the other hand, Raveau et al. found that doping of Cr and Co induces an insulator-metal transition in the insulating antiferromagnetic phase of $Pr_{0.5}Ca_{0.5}MnO_3$ (PCMO) without applying magnetic field [9]. Similarly magnetic phase diagram of $Nd(Mn_{1-x}Cr_x)O_3$ have been proposed, in which for x>0.6, there is a transition from A-type antiferromagnetic state to the ferromagnetic state [10]. However there are only few studies on the thermal properties of Cr doped $La_{0.7}Ca_{0.3}MnO_3$ (LCMO) in ferromagnetic state, where spin ordering have crucial role. Thermal critical exponents and amplitudes near the metal insulator or magnetic phase transitions determine the type of ordering and the ensuing process [11]. Martin et al. investigated the critical exponent of $La_{0.7}Sr_{0.3}MnO_3$ (LSMO) single crystal by neutron study [12]. Subharangsh Traan et al. studied thermal properties of the LCMO, PCMO, LSMO and showed a correlation between transport and magnetic properties [13]. There is a large contradiction in physical properties and thermal exponent values of CMR materials from sample to sample. The lack of data for complete phase diagram and the large variation in the physical property results of manganite materials motivated us to study the effect of $Cr^{+3}$ ion doping in $La_{0.7}Ca_{0.3}MnO_3$ manganite at close intervals. In this paper we present transport, magnetic and thermal behaviour of Cr-doped LCMO manganites. We conclude from our magnetic, IR and thermal studies that Cr dilutes double-exchange (DE) based FM of LCMO and rather promotes the AFM based super exchange interactions (SE) via $Cr^{3+}/Mn^{4+}$ ions.

## 2. EXPERIMENTAL

Samples with nominal composition $La_{0.7}Ca_{0.3}Mn_{1-x}Cr_xO_3$ ($0 \leq x \leq 1$) have been prepared by conventional solid state reaction method. Stoichiometric ratio of $La_2O_3$, $CaCO_3$, $Cr_2O_3$ and $MnO_2$ (all from Sigma Aldrich chemical Ltd with 99.9% purity) were mixed thoroughly to get homogeneous powders which were calcined at $1250^0C$ for 24h. Such calcined mixtures were then pressed into pellets and sintered in air at 1400$^0$C for 36h. XRD patterns were recorded at RIGAKU MINI FLEX II with Cu Kα radiation (1.54 Å). Electrical resistivity as a function of temperature was measured by conventional four probe method and IR spectra were recorded on Nicolet FTIR-5700 spectrometer. Magnetic measurements were carried out using SQUID magnetometer (Quantum Design, MPMS). The magnetization was measured under zero-field cooled (ZFC) and field cooled (FC) conditions with 0.01 T applied field. The heat capacity ($C_P$) measurements were carried out on PPMS (Physical property measurement system) from Quantum Design.

## 3. RESULTS AND DISCUSSIONS

### (i) Crystal Structure

Figure 1 exhibits the single phased x-ray patterns of $La_{0.7}Ca_{0.3}Mn_{1-x}Cr_xO_3$ ($0 \leq x \leq 1$) series powders. Using Rietveld refinement we were able to index the XRD peaks with orthorhombic structure in *Pbnm* space group. Table-1 provides the lattice parameters of $La_{0.7}Ca_{0.3}Mn_{1-x}Cr_xO_3$ ($0 \leq x \leq 1$) series. A regular shift towards higher 2θ is observed with increasing Cr concentration indicating a decrease in volume. Decreases in lattice volume with the Cr doping can be attributed to the smaller ionic size of $Cr^{+3}$ (0.62 Å) than that of $Mn^{+3}$ ion (0.64 Å) [14].

### (ii) Resistivity Measurement:

Figure 2 shows the electrical resistivity behaviour of the $La_{0.7}Ca_{0.3}Mn_{1-x}Cr_xO_3$ ($0 \leq x \leq 1$) series. An insulator-metal (I-M) transition is observed at 260*K* for the pristine $La_{0.7}Ca_{0.3}MnO_3$ sample which shifts to low temperatures with Cr-doping. With increasing Cr-content an additional broad peak is also seen. For 20% Cr-doping the high temperature I-M transition gets disappeared and only the broad peak is visible. For x>0.2, all samples depict the insulating behavior. However at x<0.10, Cr does not affect transition temperature faster.

These results can be understood in two ways; first the replacement of $Mn^{3+}$ with $Cr^{3+}$ can give rise to ferromagnetic (FM) interaction via double-exchange (DE) coupling as $Cr^{3+}$ ion has same electronic structure ($3d^3$, $t^3_{2g} e_g^0$) as $Mn^{4+}$ and can be connected to $Mn^{3+}$ ion via oxygen in a similar way as for $Mn^{4+}$. Therefore, the double-exchange interaction between $Mn^{3+}/Cr^{3+}$ and $Mn^{3+}/Mn^{4+}$ pairs may be thought responsible for the high temperature peak and a small sensitivity about transition temperature in the Cr-doped samples. The second peak may result from the anti-ferromagnetic (AFM) interaction due to super-exchange mechanisms between $Cr^{3+}/Cr^{3+}$ and/or $Cr^{3+}/Mn^{4+}$ ions. There seems to be a competition between double-exchange and super-exchange mechanisms in these compounds. Second at low doping ($x \leq 0.10$), as the $Cr^{3+}$ ion environment mainly consists of $Mn^{3+}$ ions, and the ratio of $Mn^{3+}$ ion and $Mn^{4+}$ ion is not much affected. Whereas, at higher doping concentrations there is an absolute reduction in $Mn^{3+}$ ion in comparison to both $Cr^{3+}$ and $Mn^{4+}$ ions. This increases AFM correlations between $Cr^{3+}/Mn^{4+}$ and $Cr^{3+}/Cr^{3+}$ via super-exchange (SE) interactions.

To further clarify this conjecture we analyzed the conduction mechanisms above $T_{IM}$. Above this temperature we fitted resistivity results by small range polaron model:

$$\rho = \rho_0\, T\, exp(E_\rho / k_B T) \qquad (1)$$

where $\rho_0$ is a constant, $E_\rho$ is the polaron activation energy, and $k_B$ is the Boltzmann constant [15, 16]. This model fits well in the insulating region except near $T_{IM}$ (inset of figure 2). The fitting parameter reveals that the activation energy $E_\rho$ increases (Table II) with increasing Cr doping. It is well known that polarons may arise due to strong lattice-electron interactions originating by Jahn-Teller distortion [5]. This makes a strong correlation between activation energy and lattice distortion. It is reasonably accepted that with smaller Cr ion doping, distortion may tend to localize the carriers hence lattice constant and activation energy have similar variation. Also as mentioned earlier, with increase in doping of Cr some of $Mn^{+4}$ ions are replaced by $Cr^{+3}$ and probability of hopping of electrons from $Mn^{+3}$ to $Mn^{+4}$ may become difficult, resulting in an increase of the activation energy. Besides the lattice distortion effect due to $Mn^{3+}/Cr^{3+}$ substitution, the double-exchange interaction between $Mn^{+3}/Cr^{+3}$ and $Mn^{+3}/Mn^{+4}$ pairs may not be similar in nature and hence affecting the conduction process.

Figure 3 reveals temperature coefficient of resistance (TCR calculated as (1/R) (dR/dT)) values for samples $La_{0.7}Ca_{0.3}Mn_{1-x}Cr_xO_3$ ($0 \leq x \leq 0.06$). For the parent compound TCR is 13.57%, however, it becomes nearly double (22.78%) for x = 0.02, and further decreases with increasing the Cr concentration. Such high TCR can have large potential application as infra-red bolometers near the transition temperature. Interestingly though the $T_{IM}$ is decreased with increased resistivity indicating relatively more localization effects, the TCR, i.e. measure of sharpness of the transition is improved in small Cr concentration regime. This could be attributed to ensuing morphological effects in terms of grain size etc, instead of pure electronic process as discussed above [17].

**(iii) IR Measurement:**

The electrical resistivity results are further corroborated by the IR spectroscopic measurements. The ideal cubic perovskite $ABO_3$ with space group $O^1_h$ has 15 normal modes of vibration, in which only three are IR active mode in range of 200 $cm^{-1}$ to 800 $cm^{-1}$ [18]. Any deviations from ideal cubic structure may cause change in vibration mode or IR spectra. IR spectra of $La_{0.7}Ca_{0.3}Mn_{1-x}Cr_xO_3$ (x = 0.0, 0.1, 0.3, 0.5, 0.7, 1.0) is observed in 400-750 $cm^{-1}$ region and is shown in Figure 4. The band due to in- phase stretching mode ($B_{2g}$ mode) of oxygen is observed at 590 $cm^{-1}$ for pristine compound that shifts to higher wave number (higher frequency) 631 $cm^{-1}$ as concentration of Cr increases. The band at 413 $cm^{-1}$ that is an $E_g$ symmetry mode associated with an internal bending mode of $MnO_6$ octahedra becomes softer and new additional secondary vibration bands are generated. As ionic charge on both $Mn^{3+}$ and $Cr^{3+}$ is the same, the geometrical/lattice effect might be mainly responsible for the change in spectra. The increase in B-O vibration frequency from 590 cm$^{-1}$ to 631 cm$^{-1}$ indicates strong coupling constant and hence the shorter bond lengths/decrease in lattice volume, supporting the XRD results.

Analyzing IR spectra one can conclude about the distribution of $Mn^{3+}$ and $Cr^{3+}$ ions. A symmetric band is expected if the $Mn^{3+}$ and $Cr^{3+}$ ions are distributed uniformly, if distribution is non-uniform then the band is expected non-symmetric. For all Cr concentration a broad shoulder appears at lower wave number side implying that all the samples have more character of $La_{0.7}Ca_{0.3}CrO_3$ with minor character $La_{0.7}Ca_{0.3}MnO_3$. The appearance of non-symmetric bands and the shoulder at lower frequency side with increasing Cr concentration indicate the uneven distribution of $Mn^{3+}$ and $Cr^{3+}$ ions. For $La_{0.7}Ca_{0.3}Mn_{0.5}Cr_{0.5}O_3$ the band is observed at 612.5 cm$^{-1}$. This composition also has a broad shoulder towards lower wave number side, indicating presence of pocketed $La_{0.7}Ca_{0.3}CrO_3$ and $La_{0.7}Ca_{0.3}MnO_3$. Other possibility in the increase of the electrical resistivity may be due to strong coupling between orbital and the spin degrees of freedom (magnetic polarons) that localize the carriers. The ferromagnetic (anti-ferromagnetic) clusters can shrink (increase) with increasing Cr concentration. Metallic character gets suppressed through the percolation mechanism that is induced by Cr ions encouraging anti-ferromagnetism [19]. This is particularly effective for higher Cr concentration ($x > 0.20$) samples.

### (iv) Magnetic Measurement:

To ascertain the presence of magnetic polarons we carried out magnetic measurements on some of the samples (Figure 5, 6). Like electrical resistivity behaviour, parent compound (LCMO) shows a paramagnetic (PM) to ferromagnetic (FM) transition with Curie temperature ($T_C$) ~255*K* near to that peak in electrical resistivity (~260 *K*). Below the transition temperature a pure ferromagnetic phase is observed. Separation in $M_{ZFC}$ and $M_{FC}$ data indicates the magnetic inhomogeneity in long range ferromagnetic ordering which is also clear from the M-H plots taken at 5*K* and 100*K* shown in figure 6 [20]. With increase in Cr doping other magnetic states like mixed FM and AFM are observed. The co-existence of FM and AFM states is supported by M-T and M-H curves. The long range ferromagnetic order is diluted even when a small amount of Cr is doped. For 10% Cr doping the ZFC and FC curves are very well bifurcated at low temperatures and there is a sudden drop in magnetic moment at around 40*K*. This sudden drop represents some sort of magnetic anisotropy or magnetic frustration in the sample. Also from M-H curves it can be seen that the magnetization is not saturated even upto 10*kOe* field and thus supports canted anti-ferromagnetic state. It has recently been reported by L.Capogna et al that for 15% Cr doping the saturation does not occur even at 3T. Further it was proposed that the canted anti-ferromagnetic state arises due to random distribution of $Cr^{3+}$ ions with random oriented spin [21] in Cr doped systems. On increasing the Cr concentration this drop becomes sharper with a decrease in magnetic moment. With increase in Cr doping, the SE driven AFM interactions become more prominent than the DE based FM. The AFM-FM phase transition is seen clearly for Cr = 0.50. Though still there is canted spin state behaviour generated by the competition between FM and AFM. Other indication can be found from the multiple peaks observed in d(ln$\chi$)/dT curves shown in inset of figure 6. For the parent compound there is a single peak but with Cr concentration an addition peak appears towards higher temperature. On complete replacement of Mn by Cr, only the strong AFM state with a little canting is present (Figure 7). The magnetic moment drops more rapidly with increase in Cr concentration. We tried to make a rough estimation for the prediction of magnetic moment of our system at 5*K*. The chemical compound with the chemical formula $La^{3+}_{0.7}Ca^{2+}_{0.3}(Mn_{1-x}Cr_x)^{3+}_{0.7}(Mn_{1-x}Cr_x)^{4+}_{0.3}O_3$ leads to a magnetic moments of

$$M(\mu_s) = [4(0.7 - 0.7x) - 3(0.7x) + 3(0.3 - 0.3x) + 2(0.3x)] \, \mu_B$$

$$= (3.7 - 5.2x) \mu_B \tag{2}$$

As $Mn^{3+}$, $Mn^{4+}$, $Cr^{3+}$ and $Cr^{4+}$ ions have magnetic moment ~ $4\mu_B$, $3\mu_B$, $3\mu_B$ and $2\mu_B$ respectively. The calculated and experimental values of magnetic moment per formula unit are listed in Table II and there variation with Cr concentration is shown in inset of figure 6. There is small discrepancy in them. For the parent compound calculated magnetic moment is 3.96 $\mu_B$, which is about 6% more than the calculated value. This may be attributed to oxygen deficiency due to which there may be small change in $Mn^{3+}$ ion concentration [22, 23]. This inhomogeneity is also supported by a small separation between FC and ZFC magnetization data. Increase in number of $Mn^{3+}$ ion results in an increase of the magnetic moment. The agreement between experimental and calculated values of the magnetic moment guarantees the near full oxygen stoichiometry of the studies samples. For higher doping (Cr > 0.5) this assumption is not valid, because the system moves towards the competing FM to AFM states with ensuing SG behavior, and hence much less observed moments. The series end compound is AFM as shown in figure 7.

In summary, from magnetization data it is observed that Cr doping dilutes long range order of ferromagnetism. The Super-exchange interactions between $Cr^{3+}$ / $Mn^{4+}$ and $Cr^{3+}$ / $Cr^{3+}$ give rise to antiferromagnetic ordering. The competing FM and AFM states may give rise to possible spin-glass type behavior in the Cr doped LCMO.

**(v) Specific Heat:**

The temperature dependence specific heat is shown in figure 8. For the sake of clarity each data has been shifted by 30 *J/mol K w.r.t.* parent LCMO sample. With increasing Cr concentration, the transition temperature ($T_{cp}$) decreases. Even with x = 0.06 the transition temperature drops to 208*K* from 249*K* for the parent compound. The relative change in transition temperature is only about 16% in comparison to Mn-site Ti doping that changes the $T_{cp}$ by 50% in LCMO system [24]. Relative less sensitivity of Cr doping towards $T_{cp}$ in comparison to Ti, indicates the possibility of some Cr being involved in double-exchange mechanism [25]. Anomalous part of specific heat ($\Delta C_p$) related to the peak at transition temperature was extracted by subtracting background from the total specific heat [see upper inset of figure 8]. To approximate our data for the specific heat, we first calculated entropy change ($\Delta S$) near the transition temperature by integrating $\Delta C_p/T$ versus T curves, which is found to decreases with increasing Cr concentration. The calculated value of $\Delta S$ are 0.36R, 0.21R and 0.06R for x = 0.0, 0.04 and 0.06 respectively, where R is the gas constant. The theoretical value for the parent compound is about Rln2, which is nearly double the experimental value. The results for the parent compound suggest some sort of inhomogeneity or partial canted spin in the ferromagnetic state [26]. This situation is abundant in most of polycrystalline LCMO [27]. With Cr substitution the observed entropy change ($\Delta S$) near the transition temperature decrease sharply and only a small fraction of the theoretical entropy is observed. This can be understood in terms of fast suppression of DE driven long range FM order being diluted by increasing SE driven AFM with short SG correlations resulting in less net entropy with progressive Cr doping.

Further to investigate the behavior around the transition temperature, we use the expression from fluctuation theory of phase transition [28];

$$C_p = A \times t^{-\alpha} \qquad (3)$$

Where t = 1- (T/$T_c$) for T< $T_{cp}$ and t = (T/$T_c$) - 1 for T>$T_{cp}$, is defined as the reduced temperatures. $\alpha$ is the critical exponent of specific heat and A is related to the critical amplitude. Calculated value of critical exponent and ratio of critical amplitude below and above the $T_{cp}$ are 0.22, 0.23 and 1.04 for x = 0.0 and 0.09, 0.09 and 0.99 for x = 0.04, see fitting of the experimental data in lower inset of figure 8. Interestingly, the x = 0.06 deviates from the theoretical fluctuation theory. Similarly for critical exponent $\beta$ at T< $T_{cp}$, we follow;

$$C_p = B \times t^{2\beta - 1} (T_c/T)^2 \qquad (4)$$

Here B is a constant and $\beta$ describes spin behavior in the ferromagnetic state [29]. The calculated values of critical exponent $\beta$ are 0.38 and 0.43 for x = 0.0, 0.04 samples respectively, which lies between theoretical Mean-field model ($\beta$ = 0.5) and Heisenberg and Ising model ($\beta$ = 0.37) [30]. For parent compound $\beta$ lies near the Ising model and Heisenberg model supports the presence of double-exchange. On the other hand with x = 0.04 critical exponent seems to move towards mean-field theory value that supports the co-existence of both long-range and short range magnetic ordering simultaneously. Also we calculated the critical exponent $\gamma$ that links the spontaneous magnetization (M) and inverse magnetic susceptibility ($\chi^{-1}$) above Curie temperature i.e. $\chi^{-1} = B ((T/T_c) - 1)^{-\gamma}$ [30]. Where B is the critical amplitude. Using scaling law;

$$\alpha + 2\beta + \gamma = 2 \qquad (5)$$

The value are found 1.01 and 1.05 for x = 0.0 and 0.04 samples respectively. $\gamma$ values are very nearer to mean field model and slight increase in value supports that with Cr doping the magnetic inhomogeneity increase.

In a nutshell with Cr doping, net entropy decreases along with critical exponent's $\beta$ and $\gamma$ increases. Decrease in net entropy indicates towards the suppression of long range magnetic order with Cr doping. For parent compound $\beta$ lies near the Heisenberg model, On the other hand with x = 0.04 the same moves towards mean-field theory, supporting the co-existence of both long-range and short range magnetic ordering in the Cr doped samples.

## 4. CONCLUSIONS

We have investigated the effect of Cr doping on structural, electrical, magnetic and thermal properties of $La_{0.7}Ca_{0.3}MnO_3$. By Rietveld refinement of the x-ray data, it is observed that Cr could replace Mn completely. The insulator-metal transition temperature decreases with Cr doping. At lower concentration of Cr, double-exchange between $Mn^{3+}$ ion & $Mn^{4+}$ ion is not affected and Cr seems to take part in double-exchange supported by resistivity and specific heat measurements. Both critical exponents' $\beta$ and $\gamma$ increases and moves towards mean-field theory. At higher concentration AFM interaction originates from super-exchange between $Cr^{3+}$ / $Mn^{4+}$ and $Cr^{3+}$ / $Cr^{3+}$ with reflection as an additional broad peak in the electrical resistivity. The magnetic measurements also corroborate the presence of competing FM and AFM interactions in the doped system.

## 5. ACKNOWLEDGEMENT

Authors acknowledge keen interest and encouragement provided by Prof. R.C. Budhani, Director-NPL. Neeraj Kumar further acknowledges the financial support from CSIR in terms of CSIR-SRF Fellowship to carry out research work at NPL and pursue for his Ph.D thesis.

**Figure Captions**

Figure 1. Reitveld fitted XRD patterns of $La_{0.7}Ca_{0.3}Mn_{1-x}Cr_xO_3$ (x= 0, 0.02, 0.70, 1.00).

Figure 2. Electrical resistivity variation with temperature of the $La_{0.7}Ca_{0.3}Mn_{1-x}Cr_xO_3$ ($0 \leq x \leq 0.2$).series

Figure 3. Temperature coefficient of resistance variation for the series $La_{0.7}Ca_{0.3}Mn_{1-x}Cr_xO_3$ ($0 \leq x \leq 0.06$)

Figure 4. IR spectra for the series $La_{0.7}Ca_{0.3}Mn_{1-x}Cr_xO_3$ ( x=0.0, 0.1, 0.3, 0.5, 0.7, 1.0).

Figure 5. Magnetization variation with temperature of the $La_{0.7}Ca_{0.3}Mn_{1-x}Cr_xO_3$ ($0 \leq x \leq 1.0$).series

Figure 6. Magnetization variation with applied field of the $La_{0.7}Ca_{0.3}Mn_{1-x}Cr_xO_3$ ($0 \leq x \leq 1.0$) series taken at 5$K$ and 100$K$. Inset shows variation of magnetic moment with doping concentration (Upper) and variation of $d(\ln\chi)/dT$ with temperature (Lower).

Figure 7. Magnetization variation of $La_{0.7}Ca_{0.3}CrO_3$. Inset shows the variation of Magnetization with temperature under zero field condition (Left) and magnetization variation with applied field taken at 5$K$, 100$K$ and 200$K$ (Right).

Figure 8. Temperature dependence of specific heat of $La_{0.7}Ca_{0.3}Mn_{1-x}Cr_xO_3$ (x= 0, 0.04 and 0.06). Upper inset (a) shows the temperature dependence anomalous part of total specific heat and lower inset (b) shows the fitting curves based on fluctuation theory.

Table I. Unit-cell parameters and reliability factor ($\chi^2$) for the refinements of $La_{0.7}Ca_{0.3}Mn_{1-x}Cr_xO_3$ ($0 \leq x \leq 1$) phases in *Pbnm* Space Group.

| ($L_{0.7}C_{0.3}M_{1-x}Cr_x O_3$) | a (Å) | b (Å) | c (Å) | $\chi^2$ | Volume (Å$^3$) |
|---|---|---|---|---|---|
| x = 0.00 | 5.462 | 5.478 | 7.720 | 1.44 | 230.959 |
| x = 0.02 | 5.462 | 5.478 | 7.719 | 2.29 | 230.950 |
| x = 0.04 | 5.460 | 5.476 | 7.718 | 2.44 | 230.749 |
| x = 0.06 | 5.459 | 5.476 | 7.717 | 2.33 | 230.683 |
| x = 0.08 | 5.459 | 5.475 | 7.716 | 2.26 | 230.612 |
| x = 0.10 | 5.457 | 5.475 | 7.714 | 2.37 | 230.484 |
| x = 0.20 | 5.452 | 5.472 | 7.709 | 2.32 | 229.977 |
| x = 0.30 | 5.443 | 5.467 | 7.701 | 2.64 | 229.302 |
| x = 0.40 | 5.438 | 5.460 | 7.691 | 3.27 | 228.378 |
| x = 0.50 | 5.432 | 5.459 | 7.683 | 4.13 | 227.846 |
| x = 0.70 | 5.427 | 5.454 | 7.676 | 4.00 | 227.211 |
| x = 1.00 | 5.436 | 5.455 | 7.686 | 2.84 | 227.927 |

Table II. Activation energy calculations from the high temperature insulating region above $T_{IM}$ using Eqn. (1), magnetic moment calculations using Eqn. (2)

| Composition ($L_{0.7}C_{0.3}M_{1-x}Cr_x O_3$) | Activation energy (meV) | $T_{IM}$ (K) | $M(\mu_B)$ Experimental | $M(\mu_B)$ Calculated |
|---|---|---|---|---|
| x=0.00 | 88.037 | 260 | 3.96 | 3.7 |
| x=0.02 | 97.954 | 247 | - | - |
| x=0.04 | 103.997 | 235 | - | - |
| x=0.06 | 112.124 | 231 | - | - |
| x=0.08 | 117.735 | 212 | - | - |
| x=0.10 | 123.501 | 210 | 3.59 | 3.18 |
| x=0.20 | - | - | 2.64 | 2.66 |
| x=0.30 | - | - | 1.88 | 2.14 |
| x=0.40 | - | - | 1.43 | 1.62 |
| x=0.50 | - | - | 0.80 | 1.1 |

Figure 1. Reitveld fitted XRD patterns of $La_{0.7}Ca_{0.3}Mn_{1-x}Cr_xO_3$ (x= 0, 0.02, 0.70, 1.00).

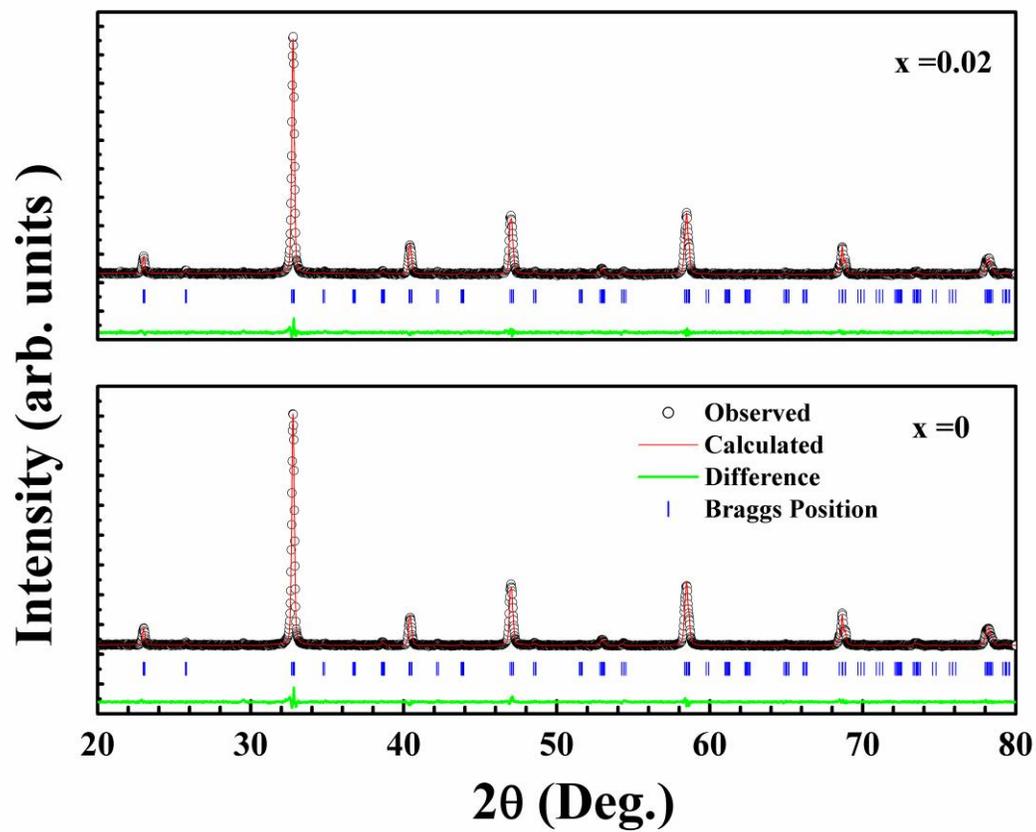

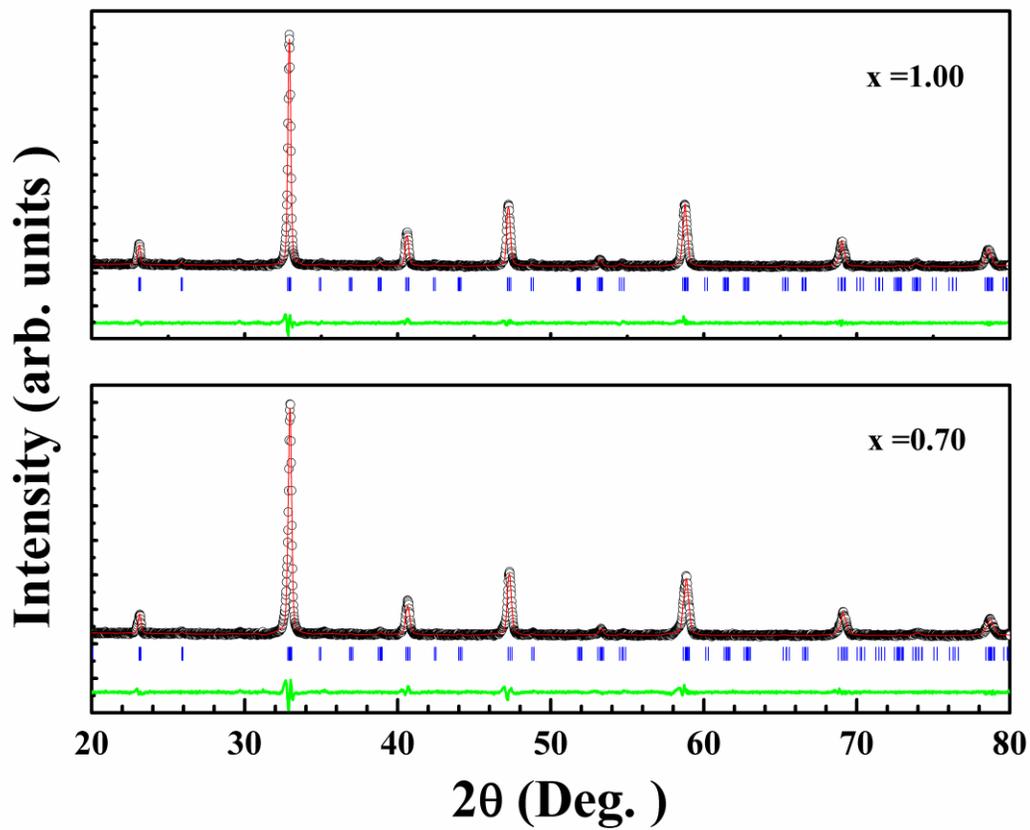

Figure 2. Electrical resistivity variation with temperature of the $La_{0.7}Ca_{0.3}Mn_{1-x}Cr_xO_3$ ($0 \leq x \leq 0.2$).series

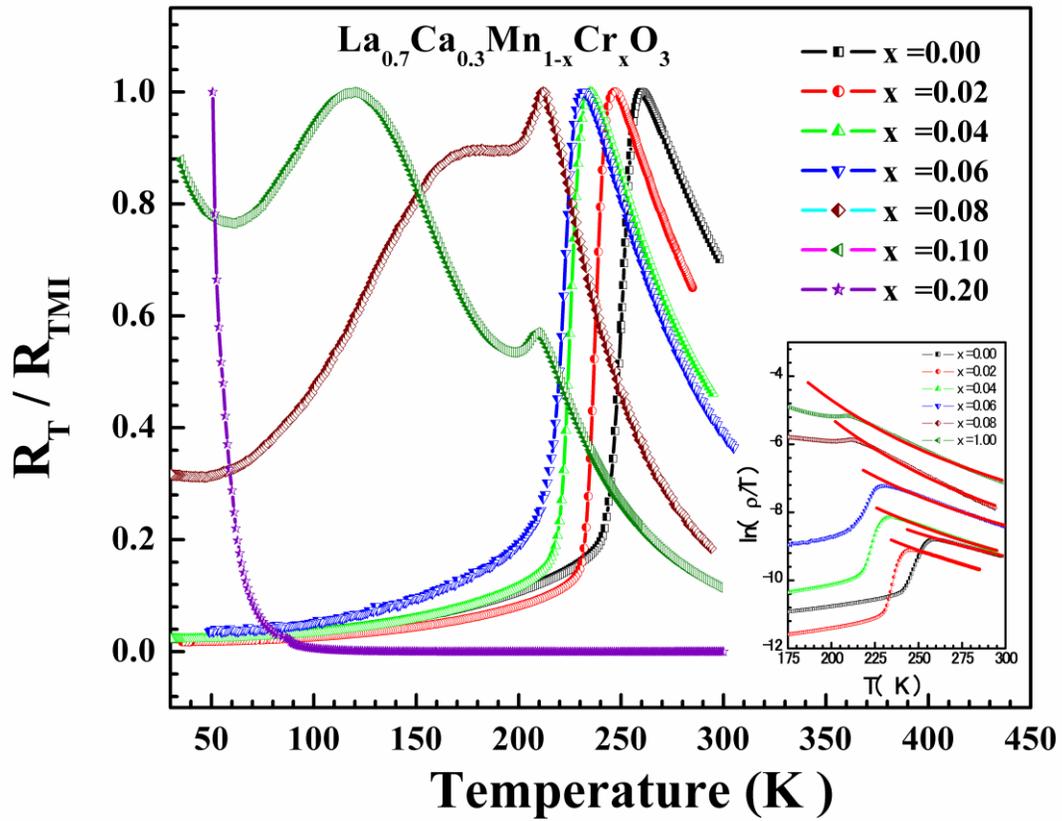

Figure 3. Temperature coefficient of resistance variation for the series $La_{0.7}Ca_{0.3}Mn_{1-x}Cr_xO_3$ ($0 \leq x \leq 0.06$)

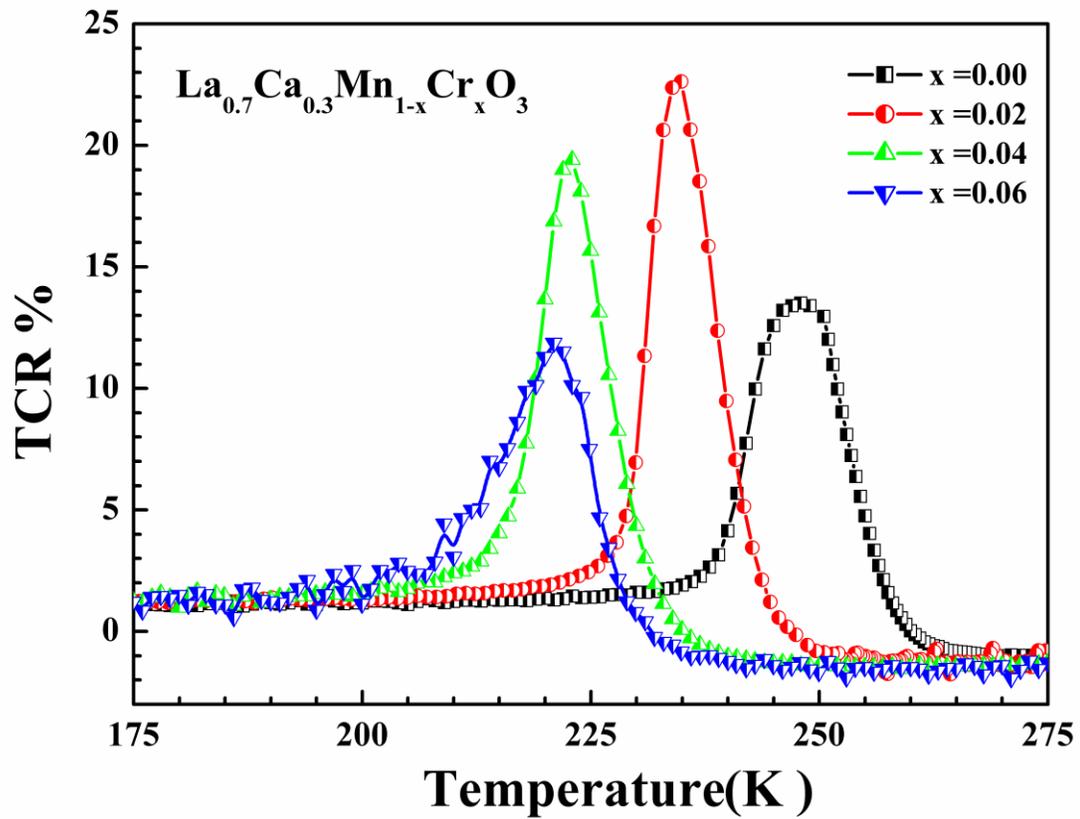

Figure 4. IR spectra for the series $La_{0.7}Ca_{0.3}Mn_{1-x}Cr_xO_3$ ( x=0.0, 0.1, 0.3, 0.5, 0.7, 1.0).

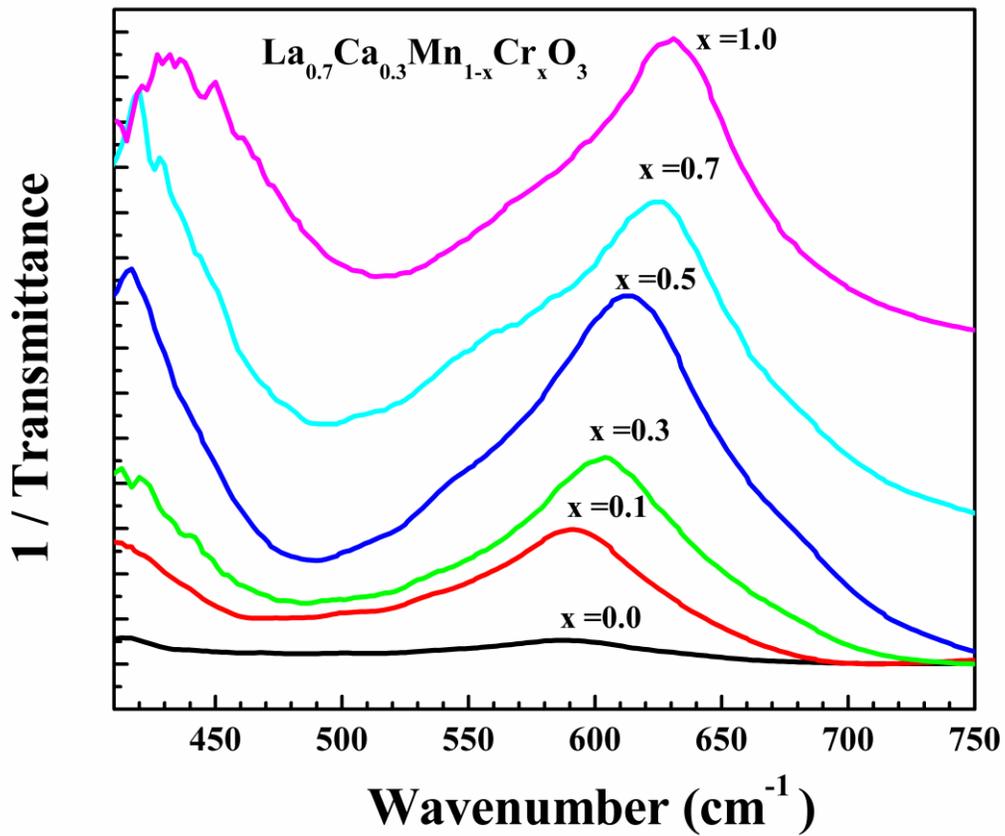

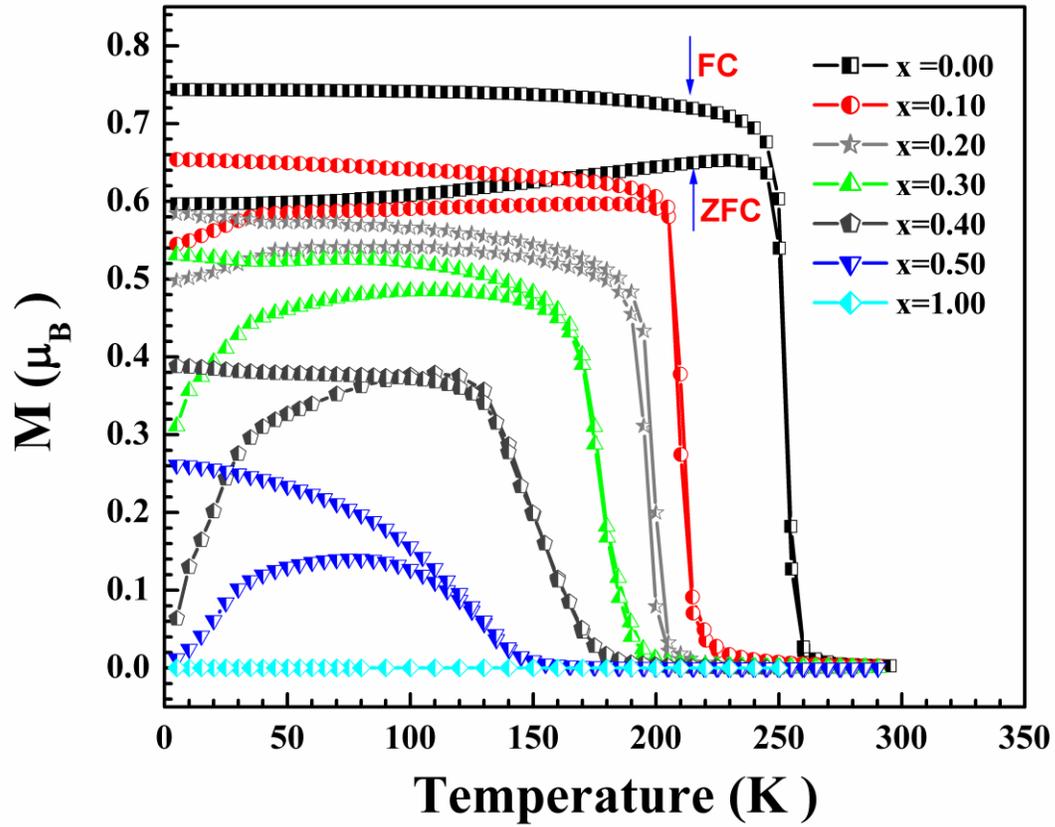

Figure 5. Magnetization variation with temperature of the $La_{0.7}Ca_{0.3}Mn_{1-x}Cr_xO_3$ ($0 \leq x \leq 1.0$).series

Figure 6. Magnetization variation with applied field of the $La_{0.7}Ca_{0.3}Mn_{1-x}Cr_xO_3$ ($0 \leq x \leq 1.0$) series taken at 5$K$ and 100$K$. Inset shows variation of magnetic moment with doping concentration (Upper) and variation of d(ln$\chi$)/dT with temperature (Lower).

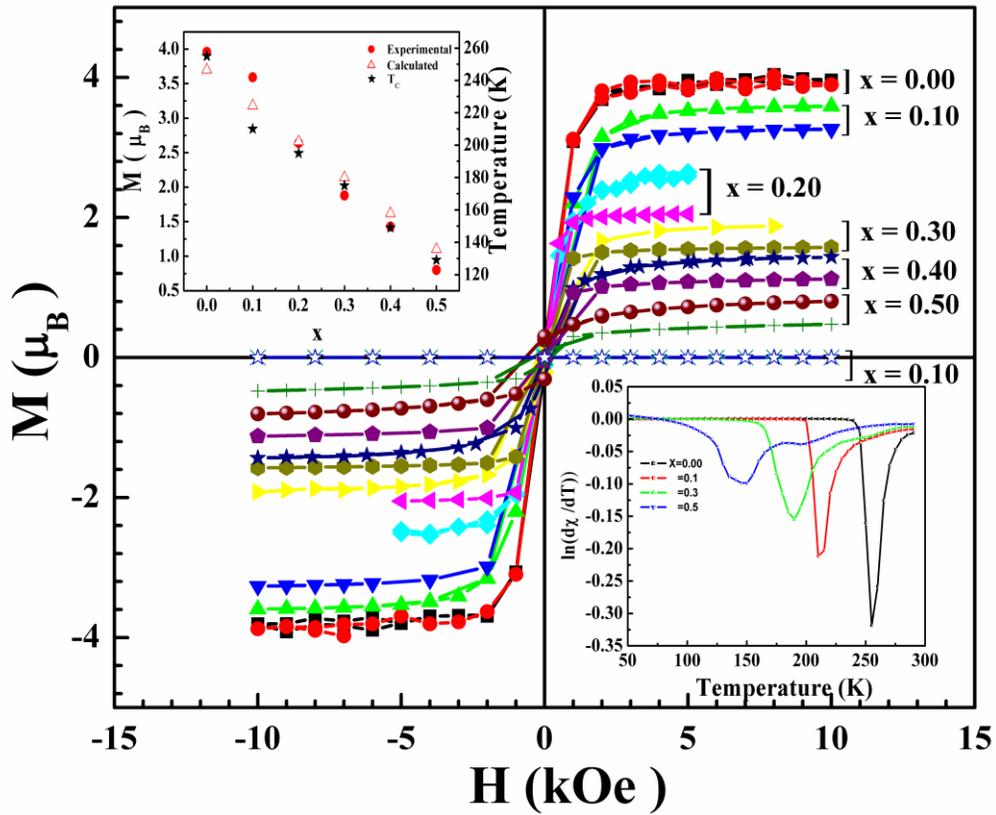

Figure 7. Magnetization variation of $La_{0.7}Ca_{0.3}CrO_3$. Inset shows the variation of Magnetization with temperature under zero field condition (Left) and magnetization variation with applied field taken at 5*K*, 100*K* and 200*K* (Right).

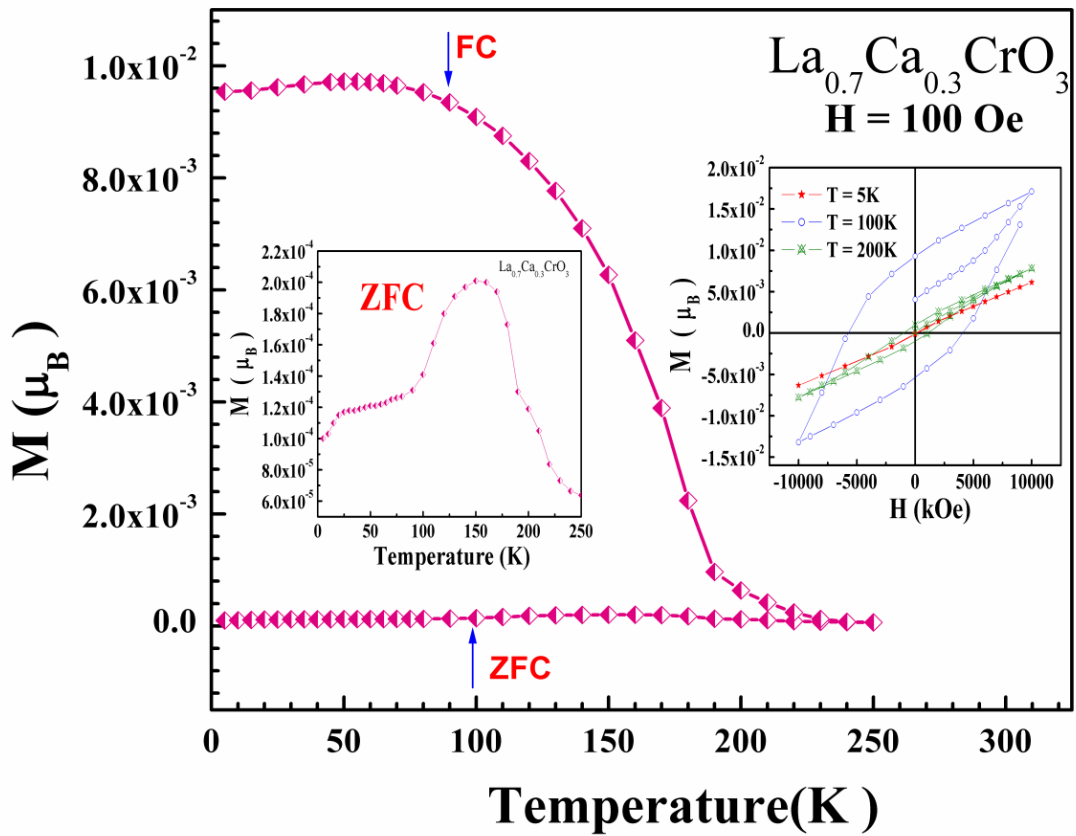

Figure 8. Temperature dependence of specific heat of $La_{0.7}Ca_{0.3}Mn_{1-x}Cr_xO_3$ (x= 0, 0.04 and 0.06). Upper inset (a) shows the temperature dependence anomalous part of total specific heat and lower inset (b) shows the fitting curves based on fluctuation theory.

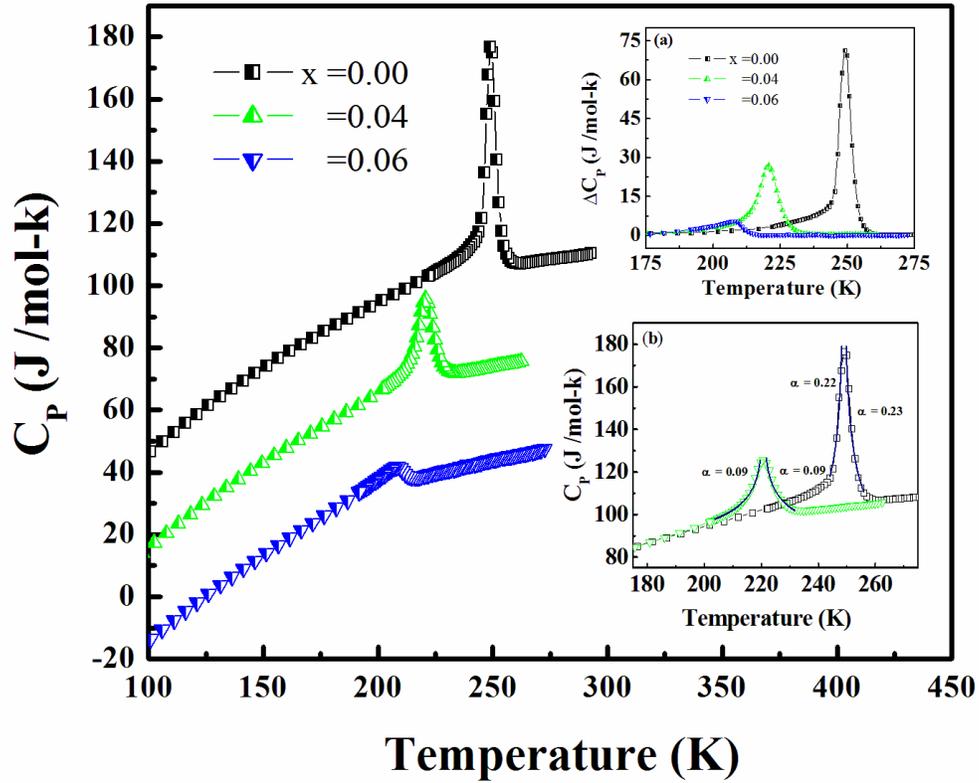